\documentclass[aps,twocolumn,longbibliography,superscriptaddress]{revtex4-2}
\usepackage{amsfonts}
\usepackage{amsmath,amssymb,epsfig}
\usepackage{float}
\usepackage{amsmath}
\usepackage{amssymb}
\usepackage{url}
\usepackage{hyperref}
\usepackage{graphicx}%
\usepackage{dcolumn}%
\usepackage{bbold}
\usepackage{siunitx}
\begin{document}
\title{Interplay of magnetic field and magnetic impurities in Ising superconductors}

\author{Wuzhang Fang}
\thanks{These two authors contributed equally to this work.}
\affiliation{Department of Physics and Astronomy and Nebraska Center for Materials and Nanoscience, University of Nebraska-Lincoln, Lincoln, Nebraska 68588, USA}
\author{M. Haim}
\thanks{These two authors contributed equally to this work.}
\affiliation{Racah Institute of Physics, Hebrew University of Jerusalem, Jerusalem 91904, Israel}
\author{K. D. Belashchenko}
\affiliation{Department of Physics and Astronomy and Nebraska Center for Materials and Nanoscience, University of Nebraska-Lincoln, Lincoln, Nebraska 68588, USA}
\author{M. Khodas}
\affiliation{Racah Institute of Physics, Hebrew University of Jerusalem, Jerusalem 91904, Israel}
\author{I. I. Mazin}
\affiliation{Department of Physics and Astronomy, George Mason University, Fairfax, VA
22030, USA}
\affiliation{Quantum Science and Engineering Center, George Mason University, Fairfax, VA
22030, USA}
\date{\today }

\begin{abstract}
Phonon-driven $s$-wave superconductivity is fundamentally antagonistic to uniform magnetism, and field-induced suppression of the
critical temperature is one of its canonical signatures. Examples of the opposite are unique and require fortuitous cancellations 
and very fine parameter tuning. The recently discovered Ising superconductors violate this rule: an external magnetic field applied in a certain direction does not suppress superconductivity in an ideal, impurity-free material. We propose a simple and experimentally accessible system where the effects of spin-conserving and spin-flip scattering can be studied in a controlled way, namely NbSe$_2$ monolayers dosed with magnetic $3d$ atoms. We predict that the critical temperature is slightly increased by an in-plane magnetic field in NbSe$_2$ dosed with Cr. Due to the band spin splitting, magnetic spin-flip scattering  requires a finite momentum transfer, while spin-conserving scattering does not. If the magnetic anisotropy is easy-axis, an in-plane field reorients the impurity spins and transforms spin-conserving scattering into spin-flip. The critical temperature is enhanced if the induced magnetization of NbSe$_2$ has a substantial long-range component, as is the case for Cr ions.
\end{abstract}
\maketitle

\section{Introduction}
It is well known that conventional superconductivity (excluding the Fulde-Ferrell-Larkin-Ovchinnikov spatially nonuniform superconductivity) cannot coexist with ferromagnetic order, but can, in principle, with ordered antiferromagnetism. This is usually rationalized in terms of superconducting coherence length being much larger that the lattice parameter, so that the staggered
magnetization averages to zero over the corresponding length scale.

This rationalization, while appealing, is doubly incorrect. First, the cancellation of the staggered magnetization is not a sufficient condition: a recently discovered class of antiferromagnets \cite{editorial,PhysRevX.12.031042} is completely compensated by symmetry, yet features a finite exchange splitting incompatible with singlet superconductivity. Neither is it necessary: another recently discovered and hotly debated phenomenon, Ising superconductivity (IS), is, as we show in this paper, compatible with ferromagnetic order provided that the magnetization direction is perpendicular to the Ising vector and the exchange field $H_{ex}$ is small compared to the spin-orbit energy splitting $\xi_{SO}$. The rationale here is that in this case the induced exchange splitting is quadratic in the exchange field as $H_{ex}^2/\xi_{SO}$, and even if the latter is much larger than the superconducting gap, $H_{ex}\gg \Delta$, the relevant parameter is still smaller: $H_{ex}^2/\xi_{SO}\ll\Delta$. 
 
IS in two-dimensional materials is a rapidly growing field of theoretical and experimental research \cite{xi2016Ising,lu2015evidence,zhou2016Ising,mockli2020Ising,sergio2018tuning,isingprx,costanzo2018tunnelling,saito2016superconductivity}. The combination of broken inversion symmetry and strong spin-orbit coupling in monolayer transition metal dichalcogenides (TMD) leads to Fermi surfaces where the electron spin is perpendicular to the plane of the monolayer, and the spin direction flips between the Fermi sheets related to each other by time reversal. This has been experimentally confirmed by establishing, for example in NbSe$_{2}$, that the
superconducting critical field is significantly higher when applied in-plane versus out-of-plane, and much larger than the Pauli limit \cite{xi2016Ising}.

Interaction of IS with impurities is highly nontrivial compared to the standard $s$-wave superconductors and has recently attracted considerable attention \cite{mockli2020Ising,nbse2_proximity,fractal}. A particularly intriguing regime is magnetic impurities with sufficiently large characteristic scattering lengths and in-plane magnetization. This regime can be realized in a material
close to ferromagnetism (superparamagnet) where a magnetic impurity generates a long-range ferromagnetic polarization. The scattering potential generated by such an impurity, in momentum space, will be strong, but also strongly peaked at small momenta. 
As we discuss below, scattering momenta smaller that the spin-orbit splitting of the Fermi contours, $\Delta k_F=\xi_{SO}/v_F$, where $v_F$ is the Fermi velocity, are not pair-breaking in the Abrikosov-Gor'kov regime \cite{AG}. 
 
This observation offers an intriguing opportunity. 
Imagine an IS that hosts magnetic defects whose magnetic moment is out-of-plane in the ground state. 
Such defects suppress superconductivity via spin-conserving scattering processes. 
As a result, the critical temperature $T_c$ exhibits the usual, almost linear decline with increasing defect concentration. 
Finally, superconductivity disappears at the critical concentration $n_{c}\sim T_{c0}/\gamma$, where $T_{c0}$ is the value of $T_c$ in the clean material. 

Suppose the defect concentration is set just above the critical one, $n\gtrsim n_{c}$, and the system is subjected to an in-plane external magnetic field $H$. 
The defect spins will then make an angle $\alpha=\cos^{-1}(HM/2\Delta E)$ with the plane, where $M$ is the magnetic moment of the defect and $\Delta E$ the magnetic anisotropy energy (MAE) per spin. 
At the saturation field $H_{s}=2\Delta E/M$ the spins complete their reorientation into the plane. 
Pair-breaking scattering off the reoriented defects has to flip the electron spin, which is initially polarized out-of-plane.
However, these scattering processes are blocked when the spatial extent of the defect magnetization exceeds $(\Delta k_F)^{-1}$.

In IS the direct pair-breaking effect of the in-plane applied magnetic field is suppressed by spin-orbit interaction. 
The same is true for the exchange fields.
Combining these statements with the blocking of the pair-breaking scattering discussed above, we conclude that the current scenario is a highly unusual example of superconductivity induced by magnetic field. 

\section{General Considerations}

Suppose that the impurity-induced polarization $m({\bf r})$ is slowly varying so that most of the spectral weight $\boldsymbol{\mu}({\bf q})=\int{d^2\mathbf{r}\,\mathbf{m}(\mathbf{r})\exp(i\mathbf{qr})}$ resides at $q\alt \Delta k_F$. Then the pair-breaking scattering rate will be reduced compared to the case of impurity spins oriented out-of-plane. 

At this point, this becomes a game of numbers, a domain of computational materials science. Indeed, if the saturation field $H_{s}$ is larger than the in-plane critical field $H_{c\parallel}$, nothing interesting happens. However, if $H_{s}<H_{c\parallel}$, then it may be possible that superconductivity at zero temperature, which is absent at zero field, will spontaneously appear at some $H_s<H < H_{c\parallel}$. Examples of superconductivity triggered by magnetic field, especially of the conventional type, are extremely rare. A canonical example is the Jaccarino-Peter effect \cite{JP}, which occurs through fine tuning of the external field to precisely compensate the existing ferromagnetic exchange field. Some heavy-fermion systems \cite{UTe2} feature \textit{reentrant} (but not newly emerging) superconductivity, which is usually interpreted in terms of triplet pairing. 

Magnetic field-triggered superconductivity in IS would, therefore, be rather unusual. By far, the best-studied IS is the NbSe$_{2}$ monolayer. Theoretically, the critical field can be as large as the spin-orbit-induced spin splitting \cite{Ilic2017,Mockli2020}, while experimental values vary from sample to sample and reach up to 40 T. This makes $H_{s}=H_{c\parallel}$ at $\Delta E\sim1.2$ meV/$\mu_{B}$, and the effect should be observable, say, for $\Delta E\lesssim0.6$ meV/$\mu_{B}$ at $H\approx H_{c\parallel}/2$. Typically, MAE of $3d$ ions is smaller than that. Another piece of information is that bulk transition-metal diselenides, including NbSe$_{2},$ can be easily intercalated with $3d$ transition-metal atoms of Cr, Mn, Fe, and Co \cite{Hauser1973,Whitney1977,iavarone2008effect}. At sizable concentrations the latter order magnetically, forming interesting and nontrivial magnetic patterns. At small concentrations they behave, in the bulk, as magnetic pair breakers, as expected. Upon exfoliation, a lightly intercalated sample would create a monolayer dosed with magnetic ions. To our knowledge, this procedure has not been performed intentionally, and the superconducting properties of such dosed monolayers have not been studied.

In this paper we present a quantitative assessment of the possible response of the Ising superconductor NbSe$_{2}$ dosed with Cr, Mn, Fe, or Co to the in-plane magnetic field. 
In section \ref{sec:model} the microscopic model is formulated that allows us to present a general theory of magnetic pair-breaking in IS in section \ref{sec:Tc}. 
Section \ref{sec:fp} describes first-principles calculations of the magnetic moments and MAE for these ions, enabling the selection of a promising candidate system for the observation of field-induced superconductivity.

\section{Model formulation}
\label{sec:model}

We represent the model Hamiltonian as a sum
\begin{align}\label{eq:Hamiltonian_sum}
    H = H_0 + H_\mathrm{SOC} + H_{p} + H^s_{dis} + H^m_{dis}\, 
\end{align}
of the kinetic energy, spin-orbit coupling, pairing interaction, scalar disorder potential and the interaction of carriers with the magnetic impurities, respectively.
Below we specify each term separately.

The first term in Eq.~\eqref{eq:Hamiltonian_sum} encapsulates the model of the band structure. 
The Fermi surface of the TMD monolayer consists of the hole pockets centered at $\Gamma$ as well as a pair of pockets at $K$ and $K'$ distinguished here by the valley index, $\eta = \pm 1$, respectively. 

\begin{figure*}
\centering
\includegraphics[width=0.85\textwidth]{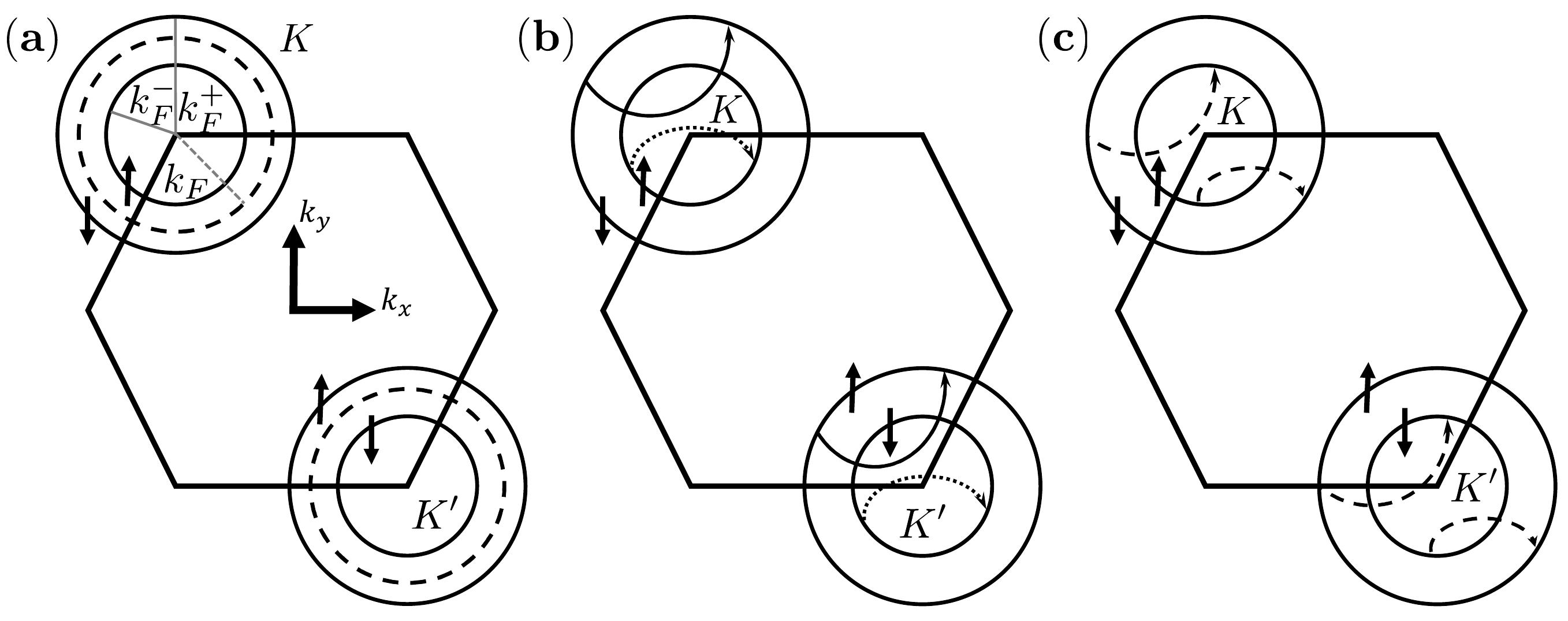}
\caption{\label{fig:K} 
In a $K$($K'$) model the two hole pockets are centered at the two corners of the hexagonal Brillouin zone. 
(a) Fermi surface spin splitting due to spin-orbit coupling. The original Fermi momentum $k_F$ for the spin-unpolarized Fermi surface turns into two distinct Fermi momenta $k_F^{\pm} = k_F \pm \xi_{\mathrm{SO}}/v_F$.
(b) Pair breaking processes active for magnetic impurities polarized out-of-plane.
(c) The same as in (b) but for in-plane polarized impurities.
}
\end{figure*}

In the monolayer, the two sets of pockets originate from different crossings of a single band with the Fermi level.
In some cases it is necessary to include both sets in the model.
In our present problem this is not essential, and we treat the two types of pockets separately, referring henceforth to the $\Gamma$ and $K$ ($K'$) models.
Although the effect of impurity spin reorientation is qualitatively similar for these two types of dispersion, it is somewhat more pronounced in the case of the $K$ ($K'$) model, which is depicted in Fig.~\ref{fig:K}. 
Indeed, spin-orbit coupling is nodeless at $K$ and $K'$ but has nodes at the Fermi surface centered at $\Gamma$. 

In Eq.~\eqref{eq:Hamiltonian_sum} we write for the dispersion
\begin{align}
H_0=\sum_{\mathbf{k},\eta,\beta}\xi_{\mathbf{k}}a_{\eta\mathbf{k}\beta}^{\dagger}a_{\eta\mathbf{k}\beta},
\end{align} 
where 
$a^\dagger_{\eta\mathbf{k}\beta}$ is the operator creating an electron in the state, $|\phi_{\mathbf{k}},\eta,\beta\rangle$ at the Fermi surface.
The states are labeled by the polar angle $\phi_{\mathbf{k}}$ formed by the momentum $\mathbf{k}$ measured from the center of the given pocket (i.e., $\Gamma$, $K$, or $K'$). 
The valley index $\eta$ is only required for the $K$ ($K'$) model.
The spin index $\beta = \pm 1$ refers to the spin up and down polarizations.
The band dispertions are approximated as $\xi_{\mathbf{k}} = k^2/2m - E_F$, where $E_F$ is the Fermi energy.
The effective mass may differ for the $\Gamma$ and $K$ ($K'$) pockets, and in both cases we have neglected the trigonal warping of the Fermi surfaces normally present in hexagonal lattices. 

Since the two types of pockets belong to the same band, the spin-orbit coupling in Eq.~\eqref{eq:Hamiltonian_sum} is given by a single function of momentum in the Brillouin zone.
Still, because of the different topology of the Fermi surfaces in the two cases, it is more convenient to formulate spin-orbit coupling separately for the two models.
For the $K$ ($K'$) model we have \cite{Xiao2012,Kormanyos2015,Ilic2017}
\begin{align}
\label{eq:Hamiltonian_SOC_K}
H_\mathrm{SOC}^K = \sum_{\eta}\underset{\mathbf{k},\beta,\beta'}{\sum}\boldsymbol{\gamma}_{\eta}\cdot\boldsymbol{\sigma}_{\beta \beta'}a_{\eta\mathbf{k}\beta}^{\dagger}a_{\eta\mathbf{k}\beta'}\, ,
\end{align}
where Ising spin-orbit coupling polarizing spins out-of-plane, $\boldsymbol{\gamma}_{\eta} =\hat{z} \gamma_{\eta}$,
is valley dependent: $\gamma_{\eta}=\eta\xi_{\mathrm{SO}}$.
In Eq.~\eqref{eq:Hamiltonian_SOC_K} and in what follows $\boldsymbol{\sigma} = (\sigma_1,\sigma_2,\sigma_3)$ stands for a vector of Pauli matrices operating in the spin space.
Spin-orbit coupling \eqref{eq:Hamiltonian_SOC_K} splits the originally spin-degenerate Fermi surface with Fermi momentum $k_F$ as shown in Fig.~\ref{fig:K}a. 
The Fermi momenta of the two spin-polarized Fermi surfaces are 
$k^{\beta\eta}_F = k_F - \eta \beta\xi_{\mathrm{SO}}/v_F$, where we have ignored the trigonal warping. 
Hence, in our notation, the electron in the state $|\phi_{\mathbf{k}},\eta,\beta \rangle$ has momentum $\mathbf{k} = k_F^{\beta\eta} \hat{\mathbf{k}}$. 

For the $\Gamma$ model we have instead of Eq.~\eqref{eq:Hamiltonian_SOC_K} \cite{Bulaevskii1976},
\begin{align}
\label{eq:Hamiltonian_SOC_G}
H_\mathrm{SOC}^\Gamma =\underset{\mathbf{k},\beta,\beta'}{\sum}\boldsymbol{\gamma}_{\mathbf{k}}\cdot\boldsymbol{\sigma}_{\beta \beta'}a_{\mathbf{k}\beta}^{\dagger}a_{\mathbf{k}\beta'}
\end{align}
where, we make an approximation $\boldsymbol{\gamma}_{\mathbf{k}} = \xi_{\mathrm{SO}}\cos(3 \phi_{\hat{\mathbf{k}}})$.

We assume the pairing interaction to be active in a spin singlet channel \cite{Smidman2017}, and in case of the $K$ ($K'$) model we write 
\begin{align}\label{eq:Hamiltonian_p}
    H_p  =   \frac{1}{2} & \sum_{\eta,\eta'}\sum_{ \beta_{1,2},\beta_{1,2}'} \underset{\mathbf{k},\mathbf{k}'}{\sum}
    v_{K}\left[i\sigma_{2}\right]_{\beta_{1}\beta_{2}}\left[i\sigma_{2}\right]_{\beta_{1}'\beta_{2}'}^{*}
    \notag \\
& \times
    a_{\eta\mathbf{k}\beta_{1}}^{\dagger}a_{\bar{\eta}\bar{\mathbf{k}}\beta_{2}}^{\dagger}
    a_{\bar{\eta}'\bar{\mathbf{k}}'\beta_{2}'}a_{\eta'\mathbf{k}'\beta_{1}'}.
\end{align}
In the $\Gamma$ model we have the pairing interaction of the same form as Eq.~\eqref{eq:Hamiltonian_p} except that the valley indices $\eta,\eta'$ are not introduced.
In both cases the interaction \eqref{eq:Hamiltonian_p} gives rise to the singlet order parameter, $\underline{\Delta}=\psi_{0}i\sigma_{2}$.

As discussed above, in this work we do not introduce the coupling between the two sets of pockets.
We note that it is essential for detailed fitting of the critical field temperature dependence \cite{Kuzmanovic2022}, in the studies of non-phonon pairing mechanisms \cite{Shaffer2020}, and is required in some scenarios with possible finite-momentum pairing \cite{Shaffer2023}.

The terms $H^s_{dis}$ and $H^{m}_{dis}$ in Eq.~\eqref{eq:Hamiltonian_sum} describe the scattering of electrons off the scalar and magnetic disorder, respectively.
In this work we disregard the scalar disorder setting $H^s_{dis}=0$. 
The effect of the scalar disorder on the phase diagram has been studied in the $\Gamma$ model, \cite{Bulaevskii1976}, in the $K$ ($K'$) model, \cite{Ilic2017}, and in the models with both types of pockets \cite{Kuzmanovic2022}.
In the absence of magnetic field and magnetization the scalar disorder is ineffective due to the Anderson theorem.
Once the time reversal symmetry is broken it becomes pair-breaking.
However, the pair breaking effect of $H_{dis}^s$ remains weak in the regime, $H,H_{ex} \ll \xi_{\mathrm{SO}}$ considered here \cite{Sosenko2017,Haim2022}. 

Based on the above observations we focus on the magnetic disorder, and model it in accordance with the first-principles calculation presented in Sec.~\ref{sec:fp}.
We consider a magnetic atom such as Cr or Fe placed on top of a NbSe$_2$ monolayer, which is responsible for the magnetic scattering of the conduction electrons. 
Its magnetic moment $\mathbf{M}$ induces spin-dependent potential both directly, through hybridization with the orbitals of the nearby host atoms which gives rise to the local interaction $J_{L}\mathbf{M}\boldsymbol{\sigma}$ of the Schrieffer-Wolff type,
and indirectly, via the induced magnetization of the host atoms near the impurity. 
The latter adds an exchange term $-\sum_k J\boldsymbol{\mu}_k\boldsymbol{\sigma}$, where $J$ is the Hund parameter and $\boldsymbol{\mu}_k$ are the magnetic moments of the host atoms $k$, which can be quite delocalized in space.

We now consider the magnetic atoms placed at random locations $\mathbf{R}_i$ on top of the NbSe$_2$ monolayer.
The average areal density of magnetic atoms is denoted as $n_m$.
We model the local and delocalized interaction of itinerant electrons with the magnetization caused by magnetic atoms in the form of the exchange coupling,
\begin{align}\label{eq:Hamiltonian_dis_m}
    H^m_{dis}  = & \sum_{\mathbf{R}_i,\mathbf{k},\mathbf{k}'}\! e^{i \mathbf{R}_i(\mathbf{k} - \mathbf{k}')}\!\!\! \sum_{\eta \beta,\eta' \beta'}
    \left[
    J \boldsymbol{\mu}(\mathbf{k}-\mathbf{k}') + J_{L} \mathbf{M}\right]\! \cdot \! \boldsymbol{\sigma}_{\beta \beta'}
    \notag \\
    & \times 
    a_{\eta\mathbf{k}\beta}^{\dagger}a_{\eta\mathbf{k}'\beta'}\, .
\end{align}

As is detailed in Sec.~\ref{sec:Tc}, the crucial feature of Eq.~\eqref{eq:Hamiltonian_dis_m} is that for $\mathbf{M}\parallel \hat{z}$ (and hence $\boldsymbol{\mu}\parallel \hat{z}$) the scattering caused by $H^m_{dis}$ is spin-conserving, while for $\mathbf{M}\perp \hat{z}$ (and $\boldsymbol{\mu}\perp \hat{z}$) it is spin-flipping. 

\section{Critical temperature}
\label{sec:Tc}

Here we compute the critical temperature of IS in the presence of the long range magnetic impurities with a prescribed magnetization axis. 
The direct pair-breaking effect of the in-plane magnetic field is negligible thanks to the protection due to Ising spin-orbit coupling, for $H \ll \xi_{\mathrm{SO}}$.

We consider two configurations, with impurities magnetized out-of-plane or in-plane. For short-range scattering, $T_c$ is the same for both polarizations \cite{Mockli2020}. 
Here we focus on the situation when the typical range of impurity potential, $\varrho$,
is comparable with or exceeds the length scale $v_F/\xi_{\mathrm{SO}}$ set by spin-orbit coupling, $\varrho \gtrsim v_F/\xi_{\mathrm{SO}}$.
An extension of the more familiar quasi-classical approach is needed in order to describe this case, and we develop such an extension below.

The basic input for the quasi-classical description are the scattering rates off the magnetic impurities. We write (half of) the rate of scattering on the $i$-th component of the magnetization as $\sigma_i(\mathbf{q})=\zeta_i \mathcal{P}(\mathbf{q})$, where $\mathbf{q}$ is the momentum transfer and $\mathcal{P}(0)\equiv1$. 
The parameter $\zeta_i$ characterizes the overall strength of the scattering, and $\mathcal{P}(\mathbf{q})$ is a dimensionless form-factor reflecting the delocalized character of the induced magnetization in the host. 

Applying the Fermi's golden rule, to the scattering potential, \eqref{eq:Hamiltonian_dis_m} one obtains, 
\begin{align}\label{eq:zeta}
    \sigma_i(\mathbf{q}) =  \pi n_m N_0 \left| J_{L} M_i - J \mu_i(\mathbf{q}) \right|^2
\end{align}
and $\zeta_i=\sigma_i(0)$, where $N_0$ is the density of host states at the Fermi level per spin. 
Because the orientation of the induced moments $\boldsymbol{\mu}_k$ relative to $\mathbf{M}$ is determined by the sign of $J_{L}$, the two terms in the scattering amplitude typically add up. We will discuss the case of Cr/NbSe$_2$ in Sec. \ref{sec:fp}.

For conciseness, we present the basic steps of the calculation as well as the results for the $K$ ($K'$) model in Sec.~\ref{sec:KK'}. 
Technical details of this calculation and a similar calculation for the $\Gamma$ model are relegated to App.~\ref{app:KK'} and App.~\ref{app:Gamma}, respectively.
The reason for this choice, is that the $K$ ($K'$) model allows for explicit results once the trigonal warping is neglected.
In contrast, the $\Gamma$ model is never isotropic due to the essential angular variation of the Ising spin-orbit coupling.

\subsection{Results for the $K$ ($K'$) model}
\label{sec:KK'}

In this section we present the results for the critical temperature $T^{\perp,\parallel}_{c}$ for magnetic impurities polarized out-of-plane or in-plane for the $K$ ($K'$) model illustrated in Fig.~\ref{fig:K}. 
To highlight our results, in the present section we present the simplified isotropic version of the $K$ ($K'$) model, where we treat the two pockets as isotropic, neglecting trigonal warping. In particular, we regard $N_0$ as unaffected by spin-orbit coupling. Under these simplifying assumptions, we state the closed result for $T^{\perp,\parallel}_{c}$ formulated in a way that is similar to, and yet distinct from the Abrikosov-Gor'kov theory \cite{AG}. 
The details of the calculations for a more general anisotropic model are relegated to the App.~\ref{app:KK'}.

In addition, we disregard intervalley scattering for the following reasons. 
First, the states at $K$ and $K'$ have orthogonal orbital characters: $d_{xy} + i d_{x^2 - y^2}$ and $d_{xy} - i d_{x^2 - y^2}$, respectively. Therefore, scattering by a $C_3$-symmetric impurity potential between the corresponding valleys is strongly suppressed \cite{Mockli2018}. Furthermore, even moderate intervalley scattering does not change the results qualitatively. Indeed, the key feature for the proposed effect is the absence of spin-flip scattering with small momentum transfer. This feature is unaffected by the inclusion of scattering to other distant regions in the Brillouin zone. This conclusion is confirmed by the calculation of the critical temperature for the $\Gamma$ model in App.~\ref{app:Gamma}.

Adopting  the discussion around Eq.~\eqref{eq:zeta} in Sec.~\ref{sec:Tc} to the $K$ ($K'$) model, we write the rate of scattering of an electron off the magnetic impurity from the initial state $|\phi_{\mathbf{k}},\eta,\beta \rangle$ into the final state $|\phi_{\mathbf{k}'},\eta',\beta'\rangle$ as
\begin{align}\label{eq:W_rate}
& W^{\eta,\eta'}_{\mathbf{k}\beta,\mathbf{k}'\beta'} =
\delta_{\eta,\eta'}  \sum_{i=1}^3 \zeta_i \mathcal{P}(\mathbf{q}_{\beta\beta'}^{\eta})
|\sigma_{\beta\beta'}^i|^2\, , 
\end{align}
where, since the scattering occurs at the Fermi surface, $\mathbf{k} = k_F^{\beta\eta} \hat{\mathbf{k}}$, the scattering momentum $\mathbf{q}_{\beta\beta'}^{\eta}$ depends on the valley and spin indices:
\begin{align}
    \mathbf{q}_{\beta\beta'}^{\eta} =
    k_F^{\beta\eta} \hat{\mathbf{k}} - k_F^{\beta'\eta} \hat{\mathbf{k}}'.
\end{align}

Let us introduce valley- and spin-dependent total scattering-out rates, 
\begin{align}\label{eq:G_rates}
    \Gamma^{\eta}_{\beta\beta'} = 
    \int_0^{2\pi} \frac{d \phi_{\mathbf{k}'}}{2\pi}
    W^{\eta,\eta}_{\mathbf{k}\beta,\mathbf{k}'\beta'}\, .
\end{align}
The rates \eqref{eq:G_rates} are not independent. 
There are two independent spin-conserving scattering rates, $\Gamma_{\beta} = \Gamma^{\eta=1}_{\beta \beta}$. The spin-conserving scattering rates for the other valley are related to $\Gamma_{\beta}$ by time-reversal symmetry: $\Gamma^{\eta}_{\beta \beta} = \Gamma^{\bar{\eta}}_{\bar{\beta} \bar{\beta}}$, where we denote $\bar{\beta} = - \beta$, and $\bar{\eta} = - \eta$. All the spin-flip scattering rates are equal thanks to the detailed balance condition for intravalley scattering: $\Gamma^{\eta}_{\beta \bar{\beta}} = \Gamma^{\eta}_{\bar{\beta} \beta }$, combined with time-reversal symmetry.
We denote the spin-flip scattering rate as $\Gamma_f$.

We express the pair-breaking effect for the out-of-plane or in-plane impurity polarization via the scattering rates introduced above:
\begin{subequations}\label{eq:AG}
\begin{equation}
\label{eq:AG_a}
\ln\left(\frac{T_{c0}}{T^{\perp}_{c}}\right)=
\frac{1}{2}\sum_{\beta}
\psi\left(\frac{1}{2}+\frac{\Gamma_{\beta}}{\pi T^{\perp}_{c}}\right)-\psi\left(\frac{1}{2}\right),
\end{equation}
\begin{equation}
\label{eq:AG_b}
\ln\left(\frac{T_{c0}}{T^{\parallel}_{c}}\right)=
\psi\left(\frac{1}{2}+\frac{\Gamma_f}{\pi T^{\parallel}_{c}}\right)-\psi\left(\frac{1}{2}\right).
\end{equation}
\end{subequations}
Here $\psi$ is the digamma function.
These results reflect the fact that the impurities polarized out-of-plane conserve, and in-plane flip, the electronic spin. 

In a standard situation, without spin-orbit coupling and/or with a short-range scattering potential the pair breaking is isotropic, we have $\Gamma_f \approx \Gamma_\beta$, and hence the spin orientation of impurities plays no role in the $T_c$ suppression.
In the present case, because of spin-orbit splitting, spin-flip scattering requires a finite momentum transfer, $|\mathbf{q}_{\beta\bar{\beta}}^{\eta}| \geq \xi_{\mathrm{SO}}/v_F$.
For the spin conserving transition, however, $|\mathbf{q}_{\beta\beta}^{\eta}|=k_{\beta \eta}^{F}[2(1-\hat{\mathbf{k}}\cdot\hat{\mathbf{k}}')]^{1/2} $ can be arbitrarily small.
Therefore, for $\varrho \gtrsim v_F/\xi_{\mathrm{SO}}$, 
$\Gamma^{\eta}_{\beta\beta} > \Gamma^{\eta}_{\beta\bar{\beta}}$.
This amounts to $\Gamma_{\beta} > \Gamma_{f}$ and  to $T_c^{\parallel} > T_{c}^{\perp}$.
As a result, field-induced superconductivity occurs at $T_c^{\perp} < T <T_{c}^{\parallel}$.

Note that Eq.~\eqref{eq:AG_a} contains an average of two digamma functions while usually, when more than one pair-breaking mechanisms are present, the pair-breaking parameters are instead added in the argument of a single digamma function \cite{Tinkham2004}.
This difference is due to the special feature of the spin-conserving scattering, which occurs
separately on two distinct spin-split Fermi surfaces.
We also note that formally the above results are consistent with the results of Ref.~\cite{Golubov1997}.
The latter, however, cannot be applied as is because, if viewed as a two-band problem, each band contains electrons with a definite spin making up only ``half'' of a Cooper pairs.

\section{First-principles calculations}
\label{sec:fp}

Since no experimental data are available for transition-metal ions deposited on monolayer NbSe$_{2}$, we first gauge our methodology against intercalated {\emph bulk} NbSe$_{2}$, where a limited amount of data is available. We consider a $3\times3\times1$ supercell constructed from the unit cell of bulk NbSe$_{2}$ ($a=3.44$ \AA, $c=12.55$ \AA) intercalated with one transition-metal (TM) atom (Cr, Mn, Fe, or Co) per supercell in its lowest energy position on top of a Nb atom \cite{Sarkar_2022}. The impurity concentration is thus 1.8\%. The structure is optimized using the projector-augmented wave (PAW) method \cite{Blochl_PAW} implemented in the Vienna \textit{Ab Initio} Simulation Package (VASP) \cite{VASP_ref,VASP_ref2}. The generalized gradient approximation of Perdew-Burke-Ernzerhof (PBE) \cite{perdew1996generalized} is used as the exchange-correlation functional. The atomic positions are relaxed using a $3\times3\times2$ $\Gamma$-centered $k$-point grid until the forces on all atoms are less than 0.01 eV/\AA. The monolayer structures are obtained by removing one of the NbSe$_2$ layers from the bulk structure and relaxed using the same criterion.

To speed up the calculations, the electronic structure and MAE of the resulting structures was studied using the OpenMX code utilizing a pseudo-atomic orbital basis set \cite{OMX1,OMX2}. The charge and spin densities were obtained from a self-consistent calculation without spin-orbit coupling and kept fixed in the subsequent MAE calculations. MAE is determined as the difference in band energies calculated with the magnetization aligned in-plane and out-of-plane. Positive MAE corresponds to easy-axis anisotropy. We have compared our results using OpenMX without LDA+U in monolayer NbSe$_2$ dosed with different impurities with the results using VASP. The agreement is excellent for Co and Mn. For Cr and Fe, the agreement is good for MAE but OpenMX gives 15\% larger magnetic moments. 

\begin{figure}[htb]
\includegraphics[width=0.9\columnwidth]{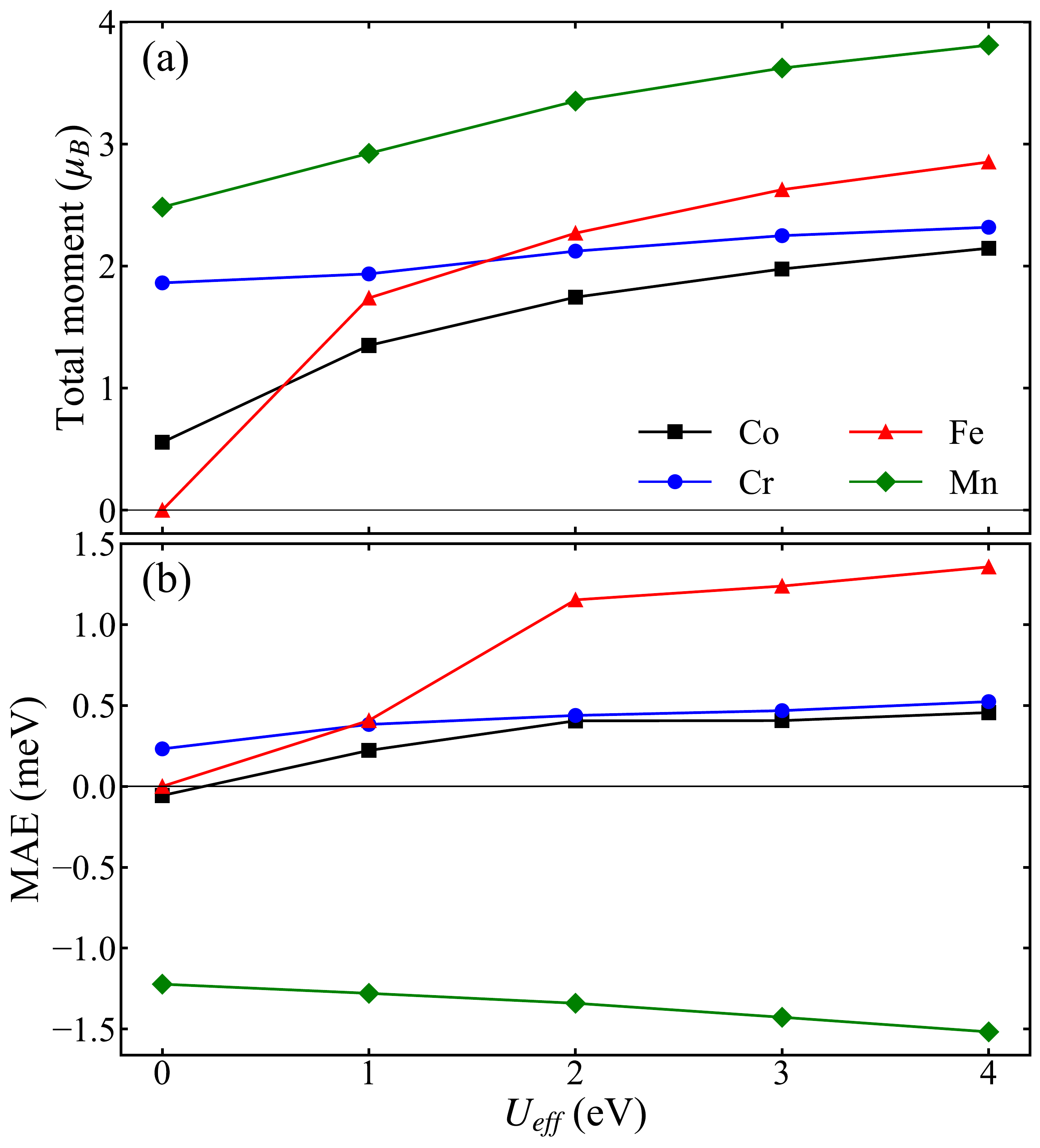}
\caption{Dependence of magnetic moments and MAEs on $U_\mathrm{eff}$ for bulk NbSe$_2$ intercalated with Cr, Mn, Fe, or Co.}
\label{fig:U-dependence}
\end{figure}

We use the LDA+U method \cite{Liechtenstein1995} to better account for the correlation effects on the $3d$ ions. Figure \ref{fig:U-dependence} shows the total magnetic moments and MAE for bulk NbSe$_2$ intercalated with different impurities as a function of $U_\mathrm{eff}=U-J$, where $U$ and $J$ are the on-site Coulomb and exchange parameters. The magnetic moments monotonically increase with $U_\mathrm{eff}$. Note that Fe-intercalated NbSe$_2$ is magnetic only if a small but finite $U_\mathrm{eff}$ is included.

The effective magnetic moments for Co$_{0.012}$NbSe$_2$, Mn$_{0.0012}$NbSe$_2$, and Fe$_{x}$NbSe$_2$ were reported at 0.6, 4.2, and 3.2 $\mu_B$, respectively \cite{iavarone2008effect,Whitney1977}. Reasonable agreement with these data for Mn and Fe can be obtained with
$U_\mathrm{eff}=4$ eV, while for Co the best agreement is obtained without LDA+U. This is in line with the general expectation that Co is the least, and Mn the most, correlated of the three ions. MAE also depends monotonically on $U$ and retains the same sign for all intercalants except Co, where it changes sign at a small value of $U_\mathrm{eff}$. Because the results for Cr (where no experimental data are available) depend little on $U_\mathrm{eff}$, we set $U_\mathrm{eff}=0$ for Cr in the following.

Having determined the optimal $U_\mathrm{eff}$ from the bulk calculations, we evaluate the MAE, the total magnetic moments, and spin-flip fields for monolayer NbSe$_2$ dosed with different impurities. The results are listed in Table \ref{tab:mae}. Here Co, Cr, and Fe have larger magnetic moments compared to the intercalated bulk, and they all have positive MAE with a spin-flip field that is less than 30 T.
For Co, the magnetic moment is close to the low-spin $3d^7$-electron configuration of Co$^{2+}$ ($S=1/2$). For Mn, it is close to the $3d^4$ configuration of Mn$^{3+}$ ($S=2$). Cr and Fe are potentially suitable candidates as they have experimentally accessible spin-flip fields.

\begin{table}[htb]
    \centering
    \begin{tabular}{|c|c|c|c|c|c|}
    \hline
    Impurity   & $U_\mathrm{eff}$ & MAE $\Delta E$ & Total moment & TM moment & $H_{s}$  \\
               & eV & meV  & $\mu_{B}$  & $\mu_{B}$  & T \\
    \hline
    Co & 0 & 1.30  & 1.57 & 1.83 & 28 \\
    \hline
    Cr & 0 & 1.44  & 3.58 & 4.31 & 14 \\
    \hline
    Fe & 4 & 1.04  & 3.47 & 3.59 & 10 \\
    \hline
    Mn & 4 & $-0.67$ & 4.56 & 4.89 & --- \\
    \hline
    \end{tabular}
    \caption{Calculated MAE, total magnetic moments, TM magnetic moments, and in-plane saturation fields in monolayer NbSe$_2$ dosed with magnetic atoms.}
    \label{tab:mae}
\end{table}

We now examine the distribution of the induced magnetic moments on the Nb atoms in Fe/NbSe$_2$ and Cr/NbSe$_2$ systems. As shown in Ref.\ \cite{Sarkar_2022}, the exchange coupling between TM adatoms on NbSe$_2$ is rather long-ranged. Therefore, we use a $15\times15$ supercell to minimize interaction between the impurities, although even at this large size the results may not be fully converged. We also note that the use of the PBE exchange-correlation potential results in a longitudinal remnant of the spin density wave (SDW) in NbSe$_2$; we suppress it by using the local density approximation (LDA) in this calculation (with $U_\mathrm{eff}$ included for Fe), which is performed in VASP.

\begin{figure}[htb]
    \centering
    \includegraphics[width=0.95\columnwidth]{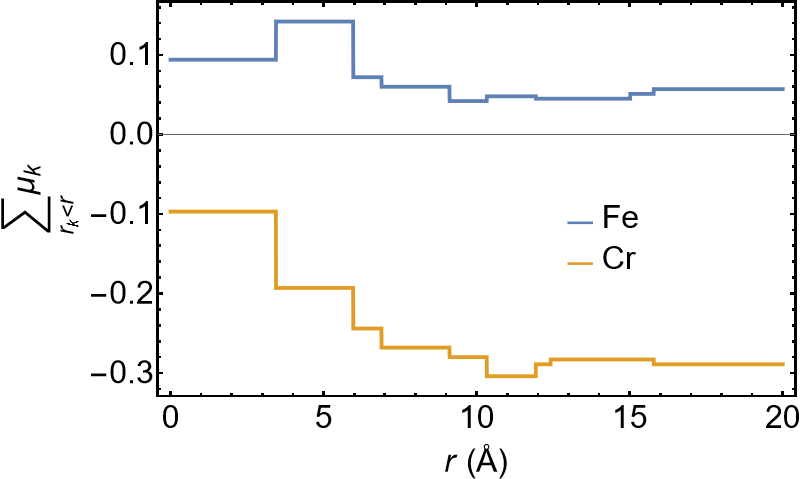}
    \caption{Total spin moment induced on Nb atoms in Fe/NbSe$_2$ and Cr/NbSe$_2$ within a circle of radius $r$ centered at the Nb atom directly underneath the impurity.}
    \label{fig:moments}
\end{figure}

Figure \ref{fig:moments} shows the total magnetic moment induced on Nb atoms within a circle of radius $r$ from the one directly underneath the impurity. In Fe/NbSe$_2$ there is a strong cancellation between positive and negative spin moments, and, as a result, the Fourier transform $\mu(\mathbf{q})$ is expected to peak at finite $\mathbf{q}$. In contrast, in Cr/NbSe$_2$ the induced magnetization is accumulated, almost monotonically, over several coordination spheres. This behavior produces $\mu(\mathbf{q})$ that is peaked at $\mathbf{q}=0$, which is favorable for field-induced superconductivity. The difference between Fe and Cr is likely due to the different hybridization patterns. In the following we focus on the Cr/NbSe$_2$ system.

According to Eq. (\ref{eq:zeta}), the scattering amplitude has contributions from local interaction with the impurity and from the delocalized interaction with the spin moments induced on the Nb atoms. The relative magnitude of these two terms can be estimated from their contributions to the exchange splitting $\Delta$ near the Fermi level in the $15\times15$ supercell. The total magnetic moment of this supercell is 3.54 $\mu_B$, which includes 5 $\mu_B$ contributed by the five filled majority-spin bands deriving from the deep $3d$ states of Cr. The remaining $-1.46$ $\mu_B$ reflect the magnetization of the native bands of NbSe$_2$. 
It includes the magnetization induced on Nb ions as well as an admixture of the minority-spin Cr orbitals.
$\Delta$ can be found by dividing this number by the DOS $N_0$ per spin in the updoped layer (1.35 eV$^{-1}$/f.u.), which gives $\Delta=4.8$ meV. Direct matching of the majority- and minority-spin DOS near the Fermi level suggests a somewhat smaller $\Delta\approx4$ meV.

\begin{figure}[htb]
    \centering
    \includegraphics[width=0.8\columnwidth]{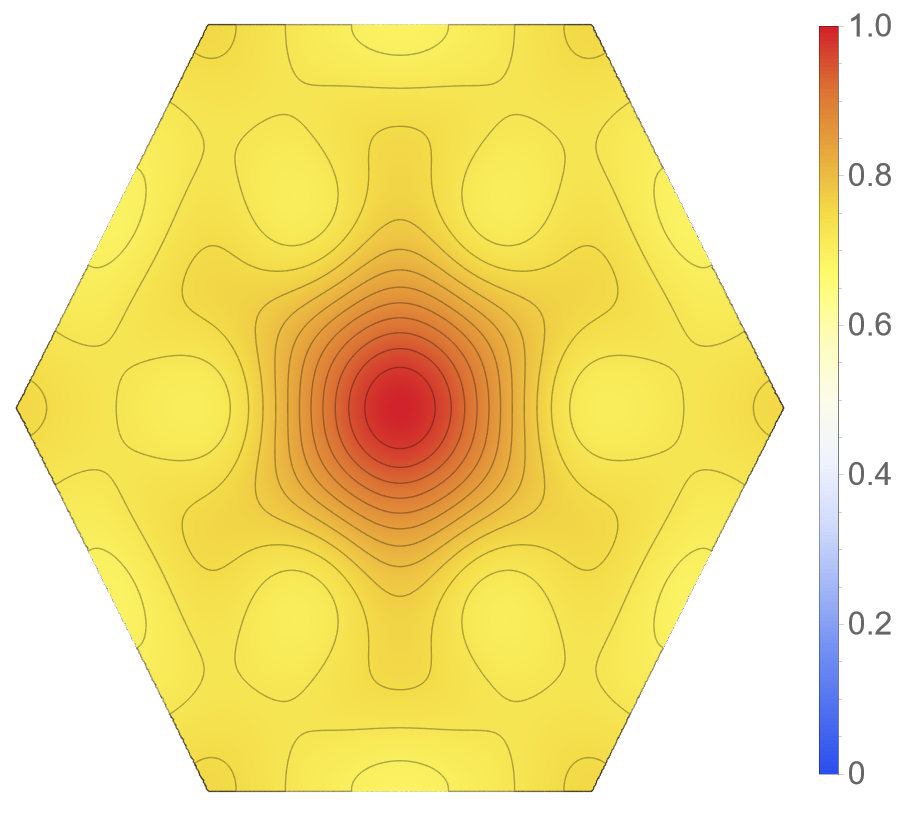}
    \caption{Scattering form-factor $\mathcal{P}(\mathbf{q})$ for Cr/NbSe$_2$ plotted in the first Brillouin zone. Contours are drawn between 0.7 and 0.975 in intervals of 0.025.}
    \label{fig:FF_full}
\end{figure}

On the other hand, the contribution of the induced Nb moments to $\Delta$ in the $15\times15$ supercell can be estimated as $J\mu_{tot}/15^2\approx 0.8$ meV, where $J\approx 0.6$ eV and $\mu_{tot}=-0.29$ $\mu_B$ is the total spin moment induced on the Nb atoms. Thus, we estimate that Hund exchange with the induced moments on the Nb atoms contributes about 20\% to the total scattering amplitude near the Fermi level, and hence $J\mu(0)\approx -0.2 J_{L}M$. (Note that $J_{L}<0$ in Cr/NbSe$_2$.)
The scattering form-factor $\mathcal{P}(\mathbf{q})$ for Cr/NbSe$_2$ is then obtained from Eq. (\ref{eq:zeta}) and shown in Fig.\ \ref{fig:FF_full}.

\section{The effect of spin reorientation}
\label{sec:comparison}

In this section we estimate the suppression of $T_c$ in NbSe$_2$ dosed with magnetic ions which occurs due to the magnetization induced in NbSe$_2$.
As discussed in section \ref{sec:Tc}, the crucial information is contained in the Fourier transform of the induced magnetization, $\mathbf{\mu}(\mathbf{q})$.

\begin{figure}[htb]
\includegraphics[width=1.0\columnwidth]{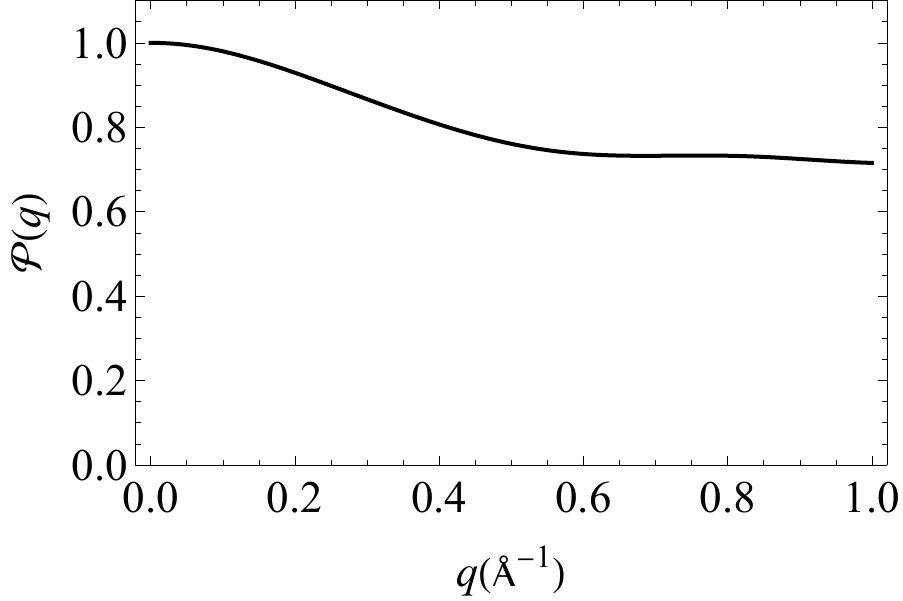}
\caption{
Isotropic approximation for the form-factor $\mathcal{P}(q)$ in Cr/NbSe$_2$ obtained by averaging $\mathcal{P}(\mathbf{q})$ shown in Fig.\ \ref{fig:FF_full} over the azimuthal angle. This approximation is applied and shown here for $q < G/2$ where $G = 4 \pi /(\sqrt{3} a)\approx 2$ \AA$^{-1}$ is the length of the shortest reciprocal lattice vector.
}
\label{fig:FormFactor}
\end{figure}

\begin{figure}[htb]
\includegraphics[width=1.0\columnwidth]{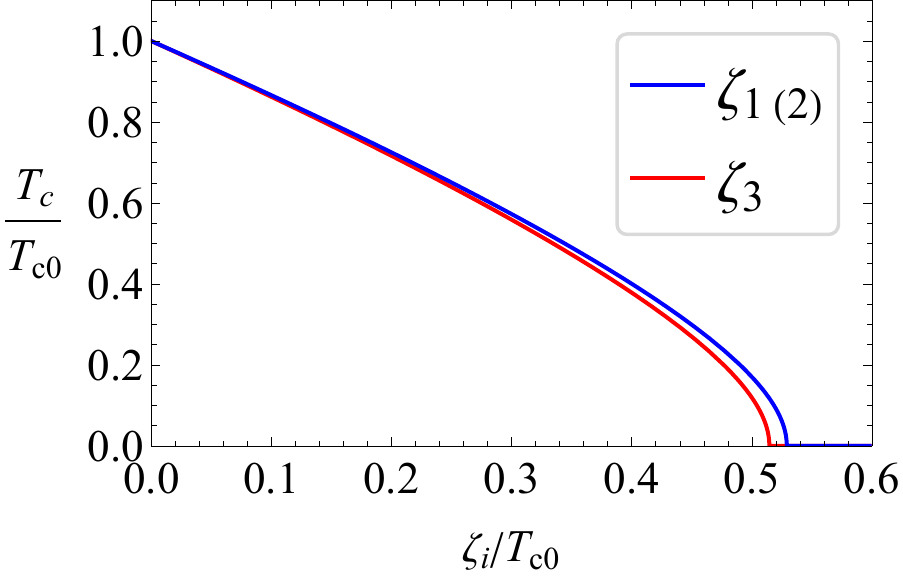}
\caption{The critical temperature $T_c$ as function of $\zeta_{i}$ with the magnetic impurities in-plane, $i=1,2$ (blue line) and out-of-plane, $i=3$ (red line), calculated by solving Eqs.~\eqref{eq:AG_b} and ~\eqref{eq:AG_a} respectively. 
The critical impurity concentration, $\zeta^{cr}_{i}$ is defined by $T_c(\zeta^{cr}_{i})=0$.
The critical concentration of impurities polarized in- and out-of-plane is $\zeta^{cr}_{1(2)} = 0.53 T_{c0}$ and $\zeta^{cr}_{3} = 0.51 T_{c0}$, respectively. 
The form factor is shown in Fig.~\ref{fig:FormFactor} and the parameters are $2mE_{F}/\hbar^{2}=0.075\AA^{-2},\ \xi_{\mathrm{SO}}/E_{F}=0.5$. $T_{c0}$ is the critical temperature without magnetic impurities}
\label{fig:TcKK}
\end{figure}

We have computed the critical temperature for the $K(K')$ model for out- and in-plane orientation of impurities.
The calculation was performed using the isotropic approximation of the form-factor shown in Fig.~\ref{fig:FormFactor}.  This approximation is sufficient because the actual momentum transfer does not exceed twice the Fermi momentum of the electronic pockets, which is about 0.7 \AA$^{-1} < G/2$. Trigonal warping can be neglected for such momenta. We have checked that this approximation produces a negligible error in the calculations of $T_c$. The results are contained in Eqs.~\eqref{eq:AG} and are presented in Fig.~\ref{fig:TcKK}.

The magnitude of the effect, namely the difference between the values of $T_c$ for two spin orientations, follows from the magnetic interaction rather localized in space. The source of this localization is the direct interaction with the magnetic ion.
We have checked that the effect is much stronger if the magnetic form-factor is less localized (see App.~\ref{app:KK'}, and Fig.~\ref{fig:TcK}). In addition, the increase in $T_c$ caused by the magnetic field occurs also in the $\Gamma$-model of the electronic dispersion. The results for the $\Gamma$-model are presented in Fig.~\ref{fig:TcGamma} of App.~\ref{app:Gamma}.

We can estimate the impurity concentration needed to achieve field-induced superconductivity as follows. For the exchange splitting $\Delta=4.8$ meV we find, using Eq.~\eqref{eq:zeta}, that the critical temperature $4$ K of a clean system is suppressed to zero at the impurity concentration of $1.4\times 10^{10}$ cm$^{-2}$. At this concentration the typical distance between the impurity atoms is on the order of 100 nm, and our analysis assuming isolated impurities is well justified.

The realization of the proposal of a field-induced superconductivity depends not only on the difference in the critical temperatures for the two spin orientations, but also on the magnitude of the parallel critical field $H_{c\parallel}(T)$ at a temperature in between the two curves shown in Fig.~\ref{fig:TcKK}. Specifically, we should have $H_{c\parallel}(T) > H_{s}$ at some point within a few percent of the critical temperature $T_c$. Although the critical field vanishes as $T_c$ goes to zero, this situation is possible because $H_{c\parallel}(T)$ rises very steeply below $T_c$. The results of Refs.~\cite{Ilic2017} and \cite{Mockli2020} show that the inequality $H_{c\parallel}(T) > H_{s}$ is satisfied even for $T_c - T = 0.02 T_c$ for non-magnetic impurities. The steep rise of $H_{c\parallel}(T)$ below $T_c$ is also expected in the case of magnetic impurities, because Ising protection is active in both cases. We have checked this numerically for our model of Cr-dozed NbSe$_2$ by solving the Gor'kov equation for $\zeta_{1(2)} = \zeta^{cr}_3$. At this concentration the critical temperature for magnetic impurities polarized out-of-plane vanishes, and that for in-plane impurities is about $0.15 T_{c0}$ (see Fig.~\ref{fig:TcKK}). We fixed the temperature at $0.1 T_{c0}$ and found $H_{c\parallel} \approx 25 T_{c0}$, which is much larger than the MAE.
Thus, the condition $H_{c\parallel}(T) > H_{s}$ may be satisfied even at a strongly suppressed critical temperature. 
In practice, other pair-breaking effects may further reduce the critical field, potentially presenting a challenge for the experimental observation of the effect of field-induced superconductivity.

It is worth noting that while the effect is sensitive to the impurity concentration --- too many impurities will kill superconductivity entirely, and too few will be not enough to have a measurable effect --- in real samples  the impurity distribution is likely to be inhomogeneous. Thus, even though the concentration may deviate from optimal in parts of the sample, there may be regions where the effect can be observed locally, for example, using electromagnetic or scanning tunneling probes.

\section{Conclusions}
\label{sec:conc}

We predicted an effect presenting another manifestation of the unique nature of Ising superconductivity: a parametric regime where 
superconductivity is strictly absent at zero external field in the presence of dilute, easy-axis magnetic impurities but can be ``turned on'' by a moderate in-plane magnetic field. Such ``incipient'' superconductivity, which requires a magnetic field to reveal itself, is unique and counterintuitive.

We presented a microscopic theory and quantitative DFT calculation of this effect, demonstrating that monolayer NbSe$_2$ lightly dosed with Cr atoms is a promising material platform for its observation. Since Cr can be easily intercalated into bulk NbSe$_2$, experimental verification of our prediction should be feasible.

The physical meaning of the effect of field-induced superconductivity is quite transparent and is similar to the famous ``Ising protection'' against the uniform in-plane magnetic field. Indeed, in the limit of dominant small-$q$ scattering (in the sense $q\alt\xi_{SO}/v_F$) the pair-breaking effect of impurities is analogous to that of Zeeman splitting; the uniform magnetic field may be viewed as a giant impurity with scattering limited to $q = 0$. 

In NbSe$_2$ the effect is relatively weak, but expected to be detectable, because of the proximity to a spin density wave at $\mathbf{Q}\approx(0.2,0,0)$\cite{Das,Das2}, which transfers some weight for small $q$ to $q\sim Q$. On the other hand, NbSe$_2$ is also close to a ferromagnetic ($q=0$) instability\cite{Das,Divilov2021}, albeit not as close as to the SDW. Potentially, an Ising superconductor that retains proximity to a ferromagnetic instability, but not to an SDW, would be an even better candidate to discover this effect. 

Another intriguing corollary comes from the fact that
the difference in $T_c$ for the two impurity spin orientations adds a new contribution to the magnetic anisotropy, which is caused by superconductivity and appears only below the superconducting transition.
Indeed, ignoring the intrinsic MAE for the moment, a strong impurity is expected to polarize spontaneously out-of-plane to avoid the loss of condensation energy within a distance of about the coherence length. This contribution to MAE is likely to exhibit a cusp-like anomaly in the temperature dependence close to the superconducting transition.
This unexpected effect will be studied elsewhere. 
Here we only note that the superconductivity-induced MAE may be reduced by the formation of Yu-Shiba-Rusinov states localized at the magnetic impurity.

However, under the assumption of weak scattering, superconductivity-induced MAE does not affect the phenomenon of field-induced superconductivity in Cr-dosed NbSe$_2$, which can be understood as follows. As assumed, weak disorder at small concentration may be treated within the Born approximation.
Spontaneous polarization is most favored if the non-flipped impurities completely destroy superconductivity.
In this situation, spontaneous polarization occurs if the condensation energy per impurity exceeds MAE.
Taking $T_{c0} \sim$ \SI{1}{\milli\electronvolt}, we can estimate condensation energy per host atom as $N_0 T_{c0}^2\sim$ \SI{1}{\micro\electronvolt}. 
This makes the condensation energy per impurity smaller than \SI{0.1}{\milli\electronvolt} at the highest impurity concentrations consistent with superconductivity for flipped impurity spins. 
As the latter is smaller than MAE by more than an order of magnitude in Cr-dosed NbSe$_2$, the spontaneous reorientation of impurity spins does not occur in the weak scattering regime. However, this effect may be stronger in other materials.

Finally, we note that the effect of field-induced superconductivity should be much more pronounced if at least one kind of band pockets is fully spin-polarized.
This case normally requires doping or working with heavier transition metals with stronger spin-orbit coupling, such as Ta.
A remarkably strong effect could be achieved if such systems were doped with magnetic atoms, assuming that MAE is strong enough to prevent spontaneous spin reorientation.

\begin{acknowledgments}
The authors acknowledge fruitful discussions with D. Wickramaratne and D. Agterberg. Work at UNL was supported by the National Science Foundation through Grant No. DMR-1916275. Support from a UNL Grand Challenges catalyst award entitled Quantum Approaches Addressing Global Threats is also acknowledged. Calculations were performed utilizing the Holland Computing Center of the University of Nebraska, which receives support from the Nebraska Research Initiative. I.I.M. was supported by the Office of Naval Research through grant N00014-23-1-2480. Some calculations were performed at the DoD Major Shared Resource Center at AFRL. M.H. and M.K. acknowledge the financial support from the Israel Science Foundation, Grant No. 2665/20.
\end{acknowledgments}

\begin{appendix}

\section{Details of the calculation of the critical temperature for the $K(K')$ model}
\label{app:KK'}
Here we present a more detailed derivation of the results, \eqref{eq:AG}. 
We have approached the current problem by extension of the quasi-classical formalism used by two of us previously to study the magnetic disorder \cite{Mockli2020}.
In the case of the $K(K')$-model we extend the usual four-dimensional space including the particle-hole (Nambu) and spin degrees of freedom to the eight dimensional space which includes in addition, the valley degree of freedom.
We denote the three sets of Pauli matrices $\boldsymbol{\rho}$, $\boldsymbol{\sigma}$ and $\boldsymbol{\tau}$ operate in Nambu, spin and valley spaces, respectively.
We also denote by $\rho_0$, $\sigma_0$ and $\tau_0$ the unit matrices operating in these spaces. 
A quantity, $\mathcal{O}$ defined in this extended eight dimensional space is denoted by the check sign, $\check{\mathcal{O}}$.
The basic object is the Green function defined as follows,  
\begin{equation}\label{eq:G}
\check{G}\left(\eta\mathbf{k},\omega_{n}\right)=\left[\begin{array}{cc}
\hat{G}^{\eta\eta}\left(\mathbf{k},\omega_{n}\right) & \hat{G}^{\eta\bar{\eta}}\left(\mathbf{k},\omega_{n}\right)\\
\hat{G}^{\bar{\eta}\eta}\left(\mathbf{k},\omega_{n}\right) & \hat{G}^{\bar{\eta}\bar{\eta}}\left(\mathbf{k},\omega_{n}\right)
\end{array}\right]\, ,
\end{equation}
where 
 $\omega_{n}=\left(2n+1\right)\pi T$ are the Matsubara frequencies with integer $n$, and
the four by four Green functions have a form,
\begin{equation}\label{eq:hat_G}
\hat{G}^{\eta\eta'}=\left[\begin{array}{cc}
G^{\eta\eta'}(\mathbf{k},\omega_{n}) & F^{\eta\eta'}(\mathbf{k},\omega_{n})\\
-[F^{\eta\eta'}(-\mathbf{k},\omega_{n})]^* & -[G^{\eta\eta'}(-\mathbf{k},\omega_{n})]^*
\end{array}\right],
\end{equation}
with each entry being a two by two matrix in spin space,
\begin{equation}
G_{\beta\beta'}^{\eta\eta'}\!(\mathbf{k},\omega_n)
\!=\!
-\!\int_0^{T^{-1}}\!\!\! d \tau e^{i \omega_n \tau}
\left\langle T_{\tau}a_{\mathbf{k}\eta \beta}\left(\tau\right)
a^{\dagger}_{\mathbf{k}\eta' \beta'}\right\rangle, 
\end{equation}
where $a_{\eta\mathbf{k}\beta}(\tau)=
e^{\tau H}a_{\eta\mathbf{k}\beta}e^{-\tau H}$. 
Within the mean field approach, the Green function, Eq.~\eqref{eq:G} satisfies the Gor'kov equation,
\begin{equation}
\label{eq:Gorkov}
\left[i\omega_{n}\check{\mathbb{1}}-\check{H}_{\mathrm{BdG}}-\check{\Sigma}\right]\check{G}=\check{\mathbb{1}}\, , 
\end{equation}
where $\check{\mathbb{1}} = \rho_0 \sigma_0 \tau_0$.
The Bogoliubov-de-Gennes (BdG) Hamiltonian takes the form 
\begin{align}
\label{eq:BdG}
&\check{H}_{\mathrm{BdG}}\!=\!\!
 \\
& \begin{bmatrix}
\xi_{k}\!+\!\boldsymbol{\gamma}_{\eta}\!\cdot\!\boldsymbol{\sigma}\! & 0 & 0 & \underline{\Delta}\\
0 & -\!\xi_{k}\!-\!\boldsymbol{\gamma}_{\eta}\!\cdot\!\boldsymbol{\sigma}^{\mathrm{T}}\! & \underline{\Delta}^{\dagger} & 0\\
0 & \underline{\Delta} & \xi_{k}\!+\!\boldsymbol{\gamma}_{\bar{\eta}}\!\cdot\!\boldsymbol{\sigma} & 0\\
\underline{\Delta}^{\dagger} & 0 & 0 & -\!\xi_{k}\!-\!\boldsymbol{\gamma}_{\bar{\eta}}\!\cdot\!\boldsymbol{\sigma}^{\mathrm{T}}
\!
\end{bmatrix}\notag
\end{align}
where for shortness we write $\xi_{k}\sigma_0$ as $\xi_{k}$, and the OP is a singlet,  $\underline{\Delta}=\psi_{0}i\sigma_{2}$.

The self-energy $\check{\Sigma}$ in the Gor'kov equation, \eqref{eq:Gorkov} incorporates the effects of scattering on the magnetic impurities:
\begin{align}
\label{eq:Self_energy_i}
\check{\Sigma}\left(\eta\mathbf{k},\omega_{n}\right)  =&
N_{0}^{-1}\pi^{-1}\sum_{i=1}^{3}\zeta_{i}\int\frac{d^{2}\mathbf{k}'}{\left(2\pi\right)^{2}}\mathcal{P}\left(\mathbf{k}-\mathbf{k}'\right)
\notag \\
& \times  
\check{V}_{i}\check{G}\left(\eta\mathbf{k}',\omega_{n}\right)\check{V}_{i}\, , 
\end{align}
where $\check{V}_1 = \rho_{3}\sigma_{1}\tau_0$, 
$\check{V}_2 = \rho_{0}\sigma_{2}\tau_0$, and 
$\check{V}_3 = \rho_{2}\sigma_{3}\tau_0$ describes the structure of the disorder scattering matrix elements due to the $i$-th component of the impurity spin polarization.
In Eq.~\eqref{eq:Self_energy_i} and throughout the paper we have neglected the scattering between the $K$ and $K'$ pockets.

The form of the BdG Hamiltonian, \eqref{eq:BdG} suggests that the dimension of the Hilbert space can be reduced.
Indeed, we show that it can be reduced to two decoupled problems in the four-dimensional space that is a product of the particle-hole and spin spaces.
The above reduction can in principle be made impossible because of the self energy, \eqref{eq:Self_energy_i}.
Yet, one can readily check by way of iterations that with this self-energy the blocks of Eq.~\eqref{eq:hat_G} satisfy,
$G^{\eta\bar{\eta}}=0$ and $F^{\eta\eta}=0$.
The matrix blocks, $\hat{G}$ take the form,
\begin{subequations}\label{eq:hat_G1}
\begin{equation}\label{eq:hat_Ga1}
\hat{G}^{\eta\eta}=\left[\begin{array}{cc}
G^{\eta\eta}(\mathbf{k},\omega_{n}) & 0\\
0 & -[G^{\eta\eta}(-\mathbf{k},\omega_{n})]^*
\end{array}\right],
\end{equation}
\begin{equation}\label{eq:hat_Ga2}
\hat{G}^{\eta\bar{\eta}}=\left[\begin{array}{cc}
0 & F^{\eta\bar{\eta}}(\mathbf{k},\omega_{n})\\
-[F^{\eta\bar{\eta'}}(-\mathbf{k},\omega_{n})]^* & 0
\end{array}\right].
\end{equation}
\end{subequations}
Equation \eqref{eq:hat_G1} shows that we can limit the consideration to the four dimensional BdG Hamiltonian comprized of the $(1,1)$, $(1,4)$,  $(4,1)$ and $(4,4)$ two by two blocks of the original BdG Hamiltonian, \eqref{eq:BdG},
\begin{align}
\label{eq:BdG1}
\hat{H}_{\mathrm{BdG}}\!=\!\!
  \begin{bmatrix}
\xi_{k}\!+\!\boldsymbol{\gamma}_{\eta}\!\cdot\!\boldsymbol{\sigma}\! &  \underline{\Delta}\\
\underline{\Delta}^{\dagger} & -\!\xi_{k}\!-\!\boldsymbol{\gamma}_{\bar{\eta}}\!\cdot\!\boldsymbol{\sigma}^{\mathrm{T}}
\end{bmatrix}
\end{align}
The same blocks of the Green function, \eqref{eq:G} according to Eq.~\eqref{eq:hat_G} comprize the reduced Green function,
\begin{align}\label{eq:hat_tilde_G}
    \hat{G}(\mathbf{k}\eta,\omega_{n}) & =\left[\begin{array}{cc}
G^{\eta\eta}(\mathbf{k},\omega_{n}) & F^{\eta\bar{\eta}}(\mathbf{k},\omega_{n})\\
-F^{\bar{\eta}\eta*}(-\mathbf{k},\omega_{n}) & -G^{\bar{\eta}\bar{\eta}*}(-\mathbf{k},\omega_{n})
\end{array}\right]
\end{align}
which has a standard form with the added information that the normal state Hamiltonian is diagonal in the valley index, while the pairing term of the Hamiltonian is off diagonal. 
The four dimensional equivalent of Eq.~\eqref{eq:Gorkov},
\begin{equation}
\label{eq:Gorkov_4}
\left[i\omega_{n}\hat{\mathbb{1}}-\hat{H}_{\mathrm{BdG}}-\hat{\Sigma}\right]\hat{G}=\hat{\mathbb{1}}\, , 
\end{equation}
where $\hat{\mathbb{1}} = \rho_0 \sigma_0$.

As is the case of the Abrikosov-Gor'kov problem, the further reduction of \eqref{eq:hat_tilde_G} down to the particle-hole space is possible.  
Specifically, we have two coupled problems defined for the inner and outer blocks of Eq.~\eqref{eq:hat_tilde_G} and labeled by the spin-index, $\beta = \pm 1$,
\begin{align}
    \bar{G}^{(\beta)}\!(\eta\mathbf{k},\omega_{n})\! =\! 
    \begin{bmatrix}\!\!
    G_{\beta\beta}^{\eta\eta}(\mathbf{k},\omega_{n}) & F_{\beta \bar{\beta}}^{\eta\bar{\eta}}(\mathbf{k},\omega_{n}) \\
    -F_{\bar{\beta} \beta}^{\bar{\eta}\eta*}(-\mathbf{k},\omega_{n}) & -G_{\bar{\beta}\bar{\beta}}^{\bar{\eta}\bar{\eta}*}(-\mathbf{k},\omega_{n})\!\!
    \end{bmatrix}
\end{align}
Let us introduce the standard definition of the quasi-classical Green function,
\begin{equation}
\label{eq:g}
\bar{g}^{(\beta)}(\eta\hat{\mathbf{k}},\omega_{n})=\int\frac{d\xi_{k}}{\pi}i\rho_{3}\bar{G}^{(\beta)}\left(\eta\mathbf{k},\omega_{n}\right).
\end{equation}
The quasi-classical Green function satisfies the Eilenberger equation,

\begin{equation}
\label{eq:Eilenberger_B_2.2}
\left[\left(i\omega_{n}\rho_{0}-\bar{S}_{\eta}^{\left(\beta\right)}-\bar{\Sigma}^{(\beta)}(\eta\hat{\mathbf{k}},\omega_{n})\right)\rho_{3},\bar{g}^{(\beta)}(\eta\hat{\mathbf{k}},\omega_{n})\right]=0
\end{equation}
where
\begin{equation}
\label{eq:S_2.2}
\bar{S}_{\eta}^{\left(\beta\right)}=\beta\left(\begin{array}{cc}
\gamma_{\eta} & \psi_{0}\\
\psi_{0}^{*} & -\gamma_{\eta}
\end{array}\right),
\end{equation}
and the self-energy follows from Eq.~\eqref{eq:Self_energy_i} by reducing it to the two-by-two matrix in the same way as we did it for the Green function.
Namely, for the contribution to the self energy from the $j$th component of the magnetization we have for the in-plane polarized impurities, $j=1,2$
\begin{equation}
\label{eq:Self_energy_j12}
\bar{\Sigma}_{j}^{(\beta)}=
-i\zeta_{j}\rho_{0}
\left\langle \bar{g}^{(\bar{\beta})}
(\eta\hat{\mathbf{k}}',\omega_{n})
\mathcal{P}( \mathbf{q}_{\beta\bar{\beta}}^{\eta})
\right\rangle _{\hat{\mathbf{k}}'}
\rho_{3},
\end{equation}
and for the impurities polarized out-of-plane we get
\begin{equation}
\label{eq:Self_energy_j3}
\bar{\Sigma}_{3}^{(\beta)}=
-i\zeta_{3}\rho_{3}\left\langle \bar{g}^{(\beta)}(\eta\hat{\mathbf{k}}',\omega_{n})
\mathcal{P}( \mathbf{q}_{\beta\beta}^{\eta})
\right\rangle _{\hat{\mathbf{k}}'}\rho_{0}.
\end{equation}
As explained above the self-energies, \eqref{eq:Self_energy_j12} and \eqref{eq:Self_energy_j3} describe spin flipping and spin conserving scattering, respectively.

To solve Eq.~\eqref{eq:Eilenberger_B_2.2} we use the following parametrization of the quasi-classical Green function,
\begin{align}\label{eq:par}
 &   \bar{g}^{(\beta)}(\eta\hat{\mathbf{k}},\omega_{n})  =i g_{0}^{(\beta)}(\eta\hat{\mathbf{k}},\omega_{n})\rho_{0}
 +g_{3}^{(\beta)}(\eta\hat{\mathbf{k}},\omega_{n})\rho_{3}
    \notag \\
&    +  if_{1}^{(\beta)}(\eta\hat{\mathbf{k}},\omega_{n})\rho_{1}
+if_{2}^{(\beta)}(\eta\hat{\mathbf{k}},\omega_{n})\rho_{2}.
\end{align}
The functions introduced in Eq.~\eqref{eq:par} satisfy the relations, 
\begin{align}
    [g_{0,3}^{(\beta)}(\eta(-\hat{\mathbf{k}}),\omega_{n})]^{*} & =
g_{0,3}^{(\bar{\beta})}(\bar{\eta}\hat{\mathbf{k}},\omega_{n}),
\notag \\
[f_{1,2}^{(\beta)}(\eta(-\hat{\mathbf{k}}),\omega_{n})]^{*} & =
f_{1,2}^{(\bar{\beta})}(\bar{\eta}\hat{\mathbf{k}},\omega_{n})\, .
\end{align}
For our purposes it is sufficient to find the functions, $f_{1,2}$ to the linear order in OP.
The linerization of the Eilenberger equation, \eqref{eq:Eilenberger_B_2.2} results in the following equations, omitting the Matsubara frequency argument for shortness, 
\begin{subequations}\label{eq:L}
\begin{align}
\label{eq:L1}
& \zeta_3 \left\langle \left( f_{1}^{(\beta)}(\eta\hat{\mathbf{k}}') +  f_{1}^{(\beta)}(\eta\hat{\mathbf{k}})\right)
\mathcal{P}(\mathbf{q}_{\beta\beta}^{\eta})
\right\rangle _{\hat{\mathbf{k}}'}
\notag \\
& -(\zeta_1 + \zeta_2) \left\langle \left( f_{1}^{(\bar{\beta})}(\eta\hat{\mathbf{k}}') - f_{1}^{(\beta)}(\eta\hat{\mathbf{k}})\right)
\mathcal{P}( \mathbf{q}_{\beta\bar{\beta}}^{\eta})
\right\rangle _{\hat{\mathbf{k}}'}
\notag \\
&  =- \left|\omega_{n}\right|f_{1}^{\left(\beta\right)}(\eta\hat{\mathbf{k}}), 
\end{align}
\begin{align}
\label{eq:L2}
& \zeta_3 \left\langle \left( f_{2}^{(\beta)}(\eta\hat{\mathbf{k}}') +  f_{2}^{(\beta)}(\eta\hat{\mathbf{k}})\right)
\mathcal{P}(\mathbf{q}_{\beta\beta}^{\eta})
\right\rangle _{\hat{\mathbf{k}}'}
\notag \\
& -(\zeta_1 + \zeta_2) \left\langle \left( f_{2}^{(\bar{\beta})}(\eta\hat{\mathbf{k}}') - f_{2}^{(\beta)}(\eta\hat{\mathbf{k}})\right)
\mathcal{P}( \mathbf{q}_{\beta\bar{\beta}}^{\eta})
\right\rangle _{\hat{\mathbf{k}}'}
\notag \\
&  =- \left|\omega_{n}\right|f_{2}^{\left(\beta\right)}(\eta\hat{\mathbf{k}}) - \beta i\psi_{0}\, .
\end{align}
\end{subequations}
Equation \eqref{eq:L1} yields $f_{1}^{(\beta)}(\eta\hat{\mathbf{k}},\omega_n)=0$.
Equation \eqref{eq:L2} is easily solved in the case of the isotropic pockets and isotropic scattering assumed here.
Thanks to the isotropy assumption, the functions   $f_{2}^{(\beta)}(\eta\hat{\mathbf{k}},\omega_n)$ are independent of the momentum direction, $\hat{\mathbf{k}}$.
Hence Eq.~\eqref{eq:L2} becomes an algebraic equation.

We consider separately the case of out- and in-plane polarized magnetic impurities.
In the case of out-of-plane impurities $\zeta_{1,2}=0$, and Eq.~\eqref{eq:L2} gives, 
\begin{equation}
\label{eq:f2_xy}
f_{2,\perp}^{(\beta)}(\eta)=-\frac{i\beta\psi_{0}}{2\Gamma^{\eta}_{\beta\bar{\beta}}+\left|\omega_{n}\right|},
\end{equation}
and for the in-plane impurities 
\begin{equation}
\label{eq:f2_z}
f_{2,\parallel}^{(\beta)}(\eta)=-\frac{i\beta\psi_{0}}{2\Gamma^{\eta}_{\beta\beta}+\left|\omega_{n}\right|}.
\end{equation}
These results has to be substituted into a linearized self-consistency conditions,
\begin{equation}
\label{eq:Self_consistency_L}
\pi T\sum_{\omega_{n},\eta}\left[\frac{\psi_{0}}{\left|\omega_{n}\right|}-\frac{i}{2}
\sum_{\beta} \beta f_{2\perp,\parallel}^{(\beta)}(\eta)\right]
=2\psi_{0}\ln\left(\frac{T_{c0}}{T_{c\perp,\parallel}}\right),
\end{equation}
The scattering rates are quadratic in the magnetization, in result, the time reversal symmetry implies 
$\Gamma^{\eta}_{\beta\beta'} = \Gamma^{\bar{\eta}}_{\bar{\beta}\bar{\beta}'}$.
We also have a detailed balance condition, $\Gamma^{\eta}_{\beta\beta'}= \Gamma^{\eta}_{\beta'\beta}$.
As a result we can eliminate the summation over the valley index in Eq.~\eqref{eq:Self_consistency_L},
\begin{equation}
\label{eq:Self_consistency_L1}
\pi T\sum_{\omega_{n}}\left[\frac{\psi_{0}}{\left|\omega_{n}\right|}-\frac{i}{2}
\sum_{\beta} \beta f_{2\perp,\parallel}^{(\beta)}(\eta=1)\right]
=\psi_{0}\ln\left(\frac{T_{c0}}{T_{c\perp,\parallel}}\right),
\end{equation}
Substituting Eqs.~\eqref{eq:f2_xy} and \eqref{eq:f2_z} into Eq.~\eqref{eq:Self_consistency_L1} we obtain Eqs.~\eqref{eq:AG}.
The summation over $\beta$ in Eq.~\eqref{eq:Self_consistency_L1} signifies that the Cooper pairs form spin singlets.
The reason we treat separately the two parts of these singlets is that the long range magnetic impurities affect them differently in the Ising superconductor.

\begin{figure}
\includegraphics[width=1.0\columnwidth]{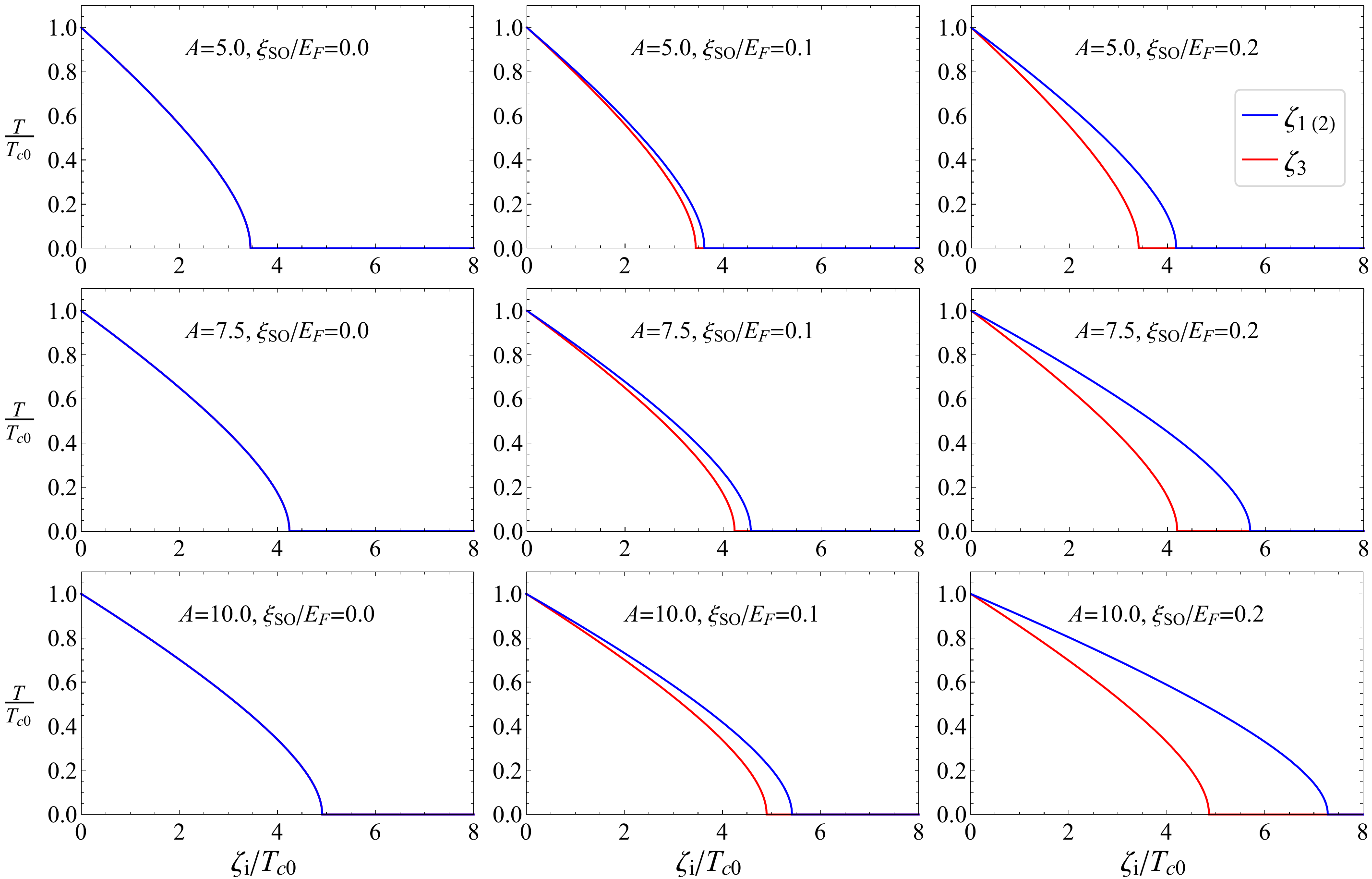}
\caption{Calculated $T_c$ as a function of the parameters, $\zeta_{i}$, $i=1,2,3$ proportional to the impurity concentration for the $K(K')$ model. 
In the calculation, only one of $\zeta_{i}$ is non-zero. For the blue lines $\zeta_{1(2)}\neq0$ and we solve  Eq.~\eqref{eq:AG_b}. For the red lines $\zeta_{3}\neq0$ and we solve Eq.~\eqref{eq:AG_a}.
Both axes are plotted in units of $T_{c0}$.
Along the rows the parameter $\xi_{SO}/E_F$ increases.
Along the columns the parameter $A = \varrho^2 m E_F/\hbar^{2}$ increases. Here we take an isotropic  Gaussian model for the form factor $\mathcal{P}\left(\mathbf{q}\right)=\exp\left[-\left(\varrho\mathbf{q}\right)^{2}/2\right]$.}
\label{fig:TcK}
\end{figure}

As discussed in the main text, the difference between the critical temperatures crucially depends on the range of the magnetic scattering.
To illustrate this point we have plotted the critical temperature for the two spin orientations for the different range of the magnetic scattering, see Fig.~\ref{fig:TcK}. 
As is clear from these plots, the more delocalized the magnetic moments are the more pronounced is the effect.

\section{The $\Gamma$ model}
\label{app:Gamma}
The $\Gamma$ pocket is singly connected and the analysis in this case is similar to the standard Abrikosov-Gor'kov theory.
The momentum dependent spin splitting is written in the form of $ \xi_{SO} \gamma(\hat{\mathbf{k}})$, and for the $\Gamma$-pocket we can take $\gamma(\hat{\mathbf{k}}) = \cos(3 \phi_{\hat{\mathbf{k}}})$.
The pair-breaking equation that controls the critical temperature, is almost identical to 
Eq.~\eqref{eq:Self_consistency_L1},
\begin{equation}
\pi T\underset{\omega_{n}}{\sum}\left[\frac{\psi_0}{\left|\omega_{n}\right|}- \frac{i}{2}\sum_{\beta }\beta \left\langle f_{2}^{(\beta)}(\hat{\mathbf{k}})\right\rangle_{\hat{\mathbf{k}}} \right]
=\psi_0 \ln\left(\frac{T_{c0}}{T_{c}}\right),
\label{eq:consistency_linear}
\end{equation}
where, as before $\langle \ldots \rangle$ stands for the averaging over the directions, ${\hat{\mathbf{k}}}$, and the index $\beta =  1(-1)$ refers to Cooper pairs with the $\uparrow \downarrow$ ($\downarrow \uparrow$), respectively.
In contrast to the $K(K')$ model the valley index is omitted. 

The two functions $f_{2}^{(\pm 1)}({\hat{\mathbf{k}}})$ satisfy the linear integral equations, 
\begin{align}
\label{eq:collision1}
& \zeta_3 \left\langle \left( f_{2}^{(\beta)}(\hat{\mathbf{k}}') +  f_{2}^{(\beta)}(\hat{\mathbf{k}})\right)
\mathcal{P}(\mathbf{q}_{\beta\beta})
\right\rangle _{\hat{\mathbf{k}}'}
\notag \\
& -(\zeta_1 + \zeta_2) \left\langle \left( f_{2}^{(\bar{\beta})}(\hat{\mathbf{k}}') - f_{2}^{(\beta)}(\hat{\mathbf{k}})\right)
\mathcal{P}( \mathbf{q}_{\beta\bar{\beta}})
\right\rangle _{\hat{\mathbf{k}}'}
\notag \\
&  =- \left|\omega_{n}\right|f_{2}^{\left(\beta\right)}(\hat{\mathbf{k}}) - \beta i\psi_{0}\, .
\end{align}
which parallels Eq.~\eqref{eq:L2}. 
As before, $\Delta\mathbf{k}_{\beta\beta'}$ is the momentum change of an electron resulting from the elastic collision off a magnetic impurity.

\begin{figure}
\includegraphics[width=1.0\columnwidth]{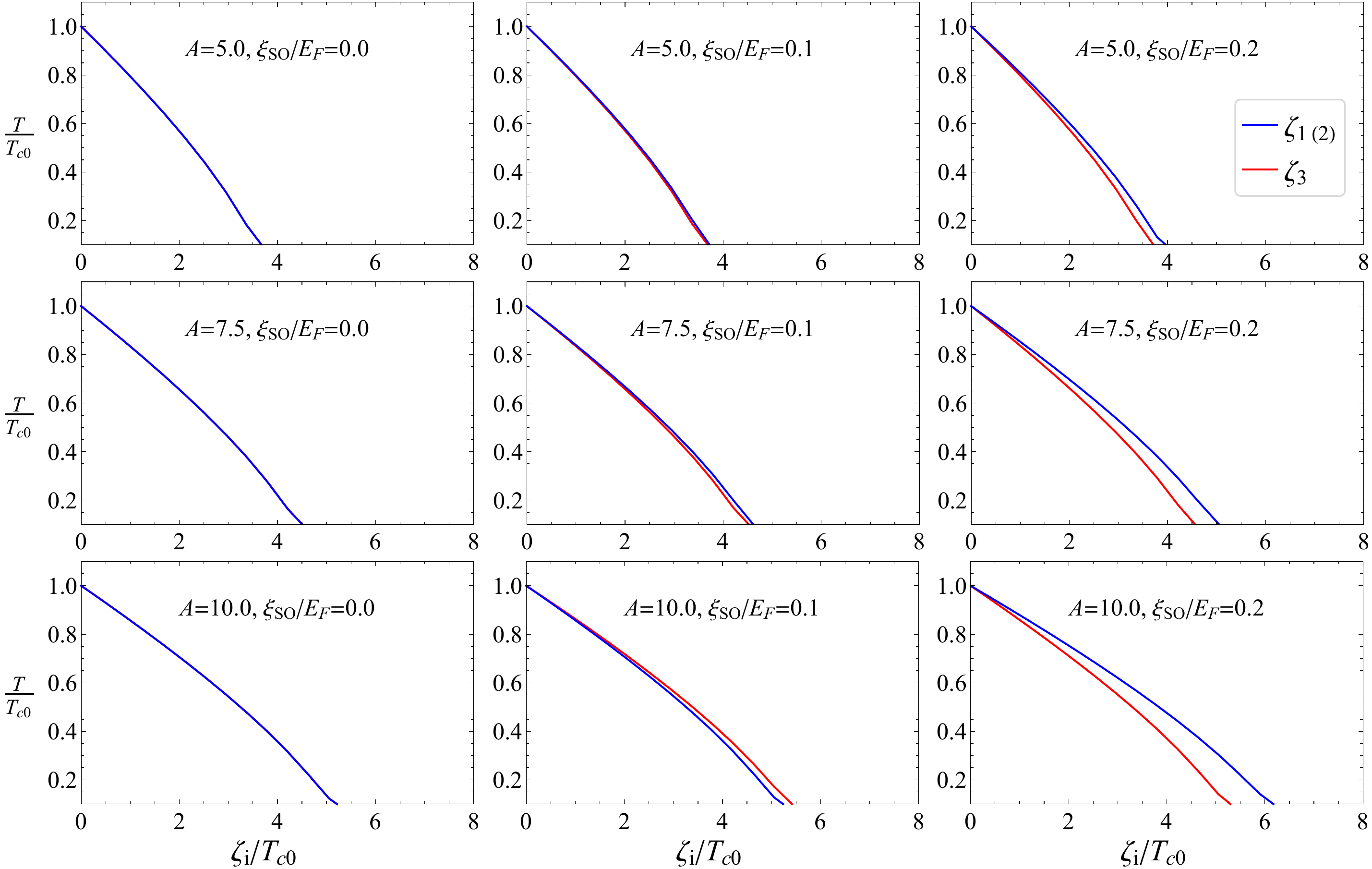}
\caption{The critical temperatures for the same set of parameters as in Fig.~\ref{fig:TcK} for the $\Gamma$ model. The calculation is done by numerically solving Eqs.~\eqref{eq:consistency_linear} and \eqref{eq:collision1} of the Fredholm type without resorting to the isotropic approximation.}
\label{fig:TcGamma}
\end{figure}

In the spin-SU(2) invariant limit, the pair-breaking effect is isotropic with respect to the polarization of magnetic impurities.
For finite spin-orbit coupling and long-range impurities it becomes anisotropic.
For the spin-conserving processes, described by the terms $\propto \zeta_3$ in Eq.~\eqref{eq:collision1}, $\mathbf{q}_{\beta\beta}$, reaches zero for $\hat{\mathbf{k}} = \hat{\mathbf{k}}'$ as the both the initial and final states belong to the same Fermi surface.
This is true for any initial momentum.

For the spin flip processes described by the terms $\propto \zeta_1,\zeta_2$, 
$\mathbf{q}_{\beta\bar{\beta}}$ remains finite and on the order of $\xi_{SO}/v_F$.
For this reason the $T_c$ is anisotropic for both $K(K')$ and $\Gamma$ models.
Since in the $\Gamma$ model the spin splitting vanishes along $\Gamma M$ directions, the transferred momentum  $\Delta\mathbf{k}_{\beta\bar{\beta}}$ becomes small for the momenta along these symmetry lines.
For this reason the $T_c$ anisotropy is more pronounced for the $K(K')$ model than for the $\Gamma$ model.

The spin-orbit coupling and, more generally, the band structure introduces anisotropy in the problem, and the constant functions of $\hat{\mathbf{k}}$ are no longer solutions of the equation \eqref{eq:collision1}.
This linear integral equation of Fredholm type can be solved numerically, and the result of the numerical solutions for a particular $\Gamma$-model are shown in Fig.~\ref{fig:TcGamma}.
These plots demonstrate that the general trend for increasing difference, $T_{c,\perp}-T_{c,\parallel}$ with the scattering range holds for the $\Gamma$ and $K$-models alike.
However, the comparison between the Figs.~\ref{fig:TcK} and \eqref{fig:TcGamma} shows that the effect for the $K$-model is stronger.
The reason for this is the vanishing of the spin-orbit coupling along $\Gamma M$ directions for the $\Gamma$-model.

\end{appendix}

\end{document}